\documentclass[twocolumn,prb,aps,floats,showpacs,amssymb,epsf,floatfix]{revtex4}
\usepackage{graphicx}
\usepackage{bm}
\usepackage{amsmath}
\usepackage{afterpage}

\begin{document}

\title{Electronic transport in normal-conductor/graphene/normal-conductor junctions and conditions for insulating behavior
 at a finite charge-carrier density}

\author{John P. Robinson and Henning Schomerus}
\affiliation{Department of Physics, Lancaster University,
Lancaster, LA1 4YB, UK}

\date{September 2007}

\begin{abstract}

We investigate the conductance of
normal-conductor/graphene/normal-conductor (NGN) junctions for
arbitrary on-site potentials in the normal and graphitic parts of
the system. We find that a ballistic NGN junction can display
insulating behavior even when the charge-carrier density in the
graphene part is finite. This effect originates in the different
${\bf k}$ intervals supporting propagating modes in graphene and a
normal conductor, and persists for moderate levels of bulk, edge
or interface disorder. The ensuing conductance thresholds could be
used as an electronic tool to map out details of the graphene band
structure in absolute ${\bf k}$-space.

\end{abstract}
\maketitle

\section{Introduction}
Graphene, the two-dimensional arrangement of carbon atoms on a
honeycomb lattice that has recently become available through
ground-breaking fabrication methods, \cite{Novoselov2004,Geim2007}
possesses a wide range of unique electronic transport properties
which originate from the conical dispersion relation around the
corners (K-points) of the hexagonal Brillouin zone.
\cite{dresselhausbook} The low-energy theory in the vicinity of
these points is of the form of a Dirac equation for massless,
chiral fermions. \cite{dirac1} The intrinsic transport properties
studied on the basis of the Dirac equation (such as the quantum
Hall effect,
\cite{Novoselov2005,Zhang2005,Novoselov2006,roomtempqhe} the
minimal conductivity,
\cite{Novoselov2005,Zhang2005,ludwig,shon,guinea1,tworzydlo,Nomura,aleiner,altland,ryu,Ostrovsky,DasSarma}
and weak localization corrections to the conductance
\cite{Morozov,Savchenko,Wu,mccann,kechedzhi,morpurgo}) therefore
effectively probe the graphene band structure via the momentum
difference $\delta {\bf k}$ relative to the K-points. On the other
hand, detailed information corroborating the conical band
structure in absolute $\bf k$ space has recently become accessible
via angle-resolved photoemission spectroscopy.
\cite{arpes1,arpes2}

\begin{figure}[tb]
\begin{center}
\includegraphics[width=\columnwidth]{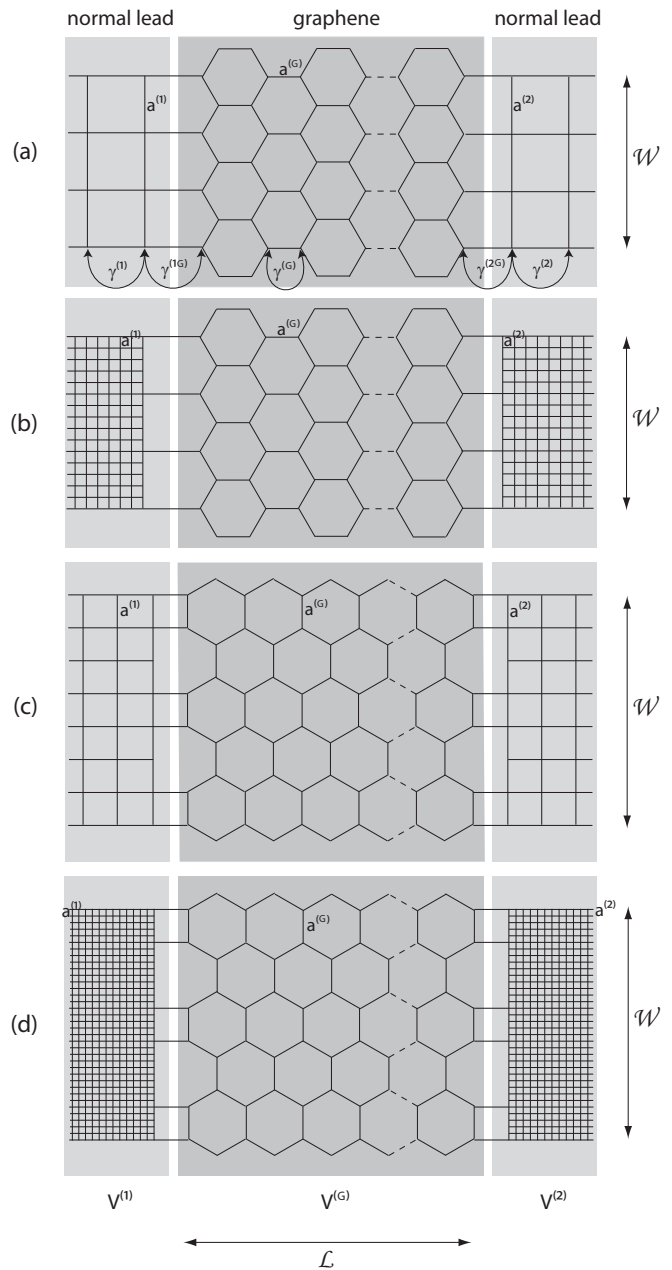}
\end{center} \caption{Tight-binding models of normal-conductor/graphene/normal-conductor (NGN) junctions, consisting
of a hexagonal lattice with lattice constant $a^{(G)}$ coupled to
square lattices with lattice constant $a^{(r)}$ ($r$=1,2
corresponding to left and right leads respectively). (a) Zigzag
interfaces connected to lattice-matched leads,
$a^{(r)}=\sqrt{3}a^{(G)}$. (b) Zigzag interfaces connected to
real-space leads, $a^{(r)}=\frac{1}{5}\sqrt{3}a^{(G)}$. (c)
Armchair interfaces connected to lattice-matched leads,
$a^{(r)}=a^{(G)}$. (d) Armchair interfaces connected to real-space
leads, $a^{(r)}=\frac{1}{5}a^{(G)}$. It is assumed that the
charge-carrier density in graphene and in the leads can be
controlled independently by gates (shaded) or chemical doping,
which shifts the on-site potentials $V^{(G)}$ (in the graphitic
region), $V^{(1)}$ (in the left lead) and $V^{(2)}$ (in the right
lead). \label{fig1}}
\end{figure}

Recent theoretical transport studies have pointed out that highly
unconventional devices could be fabricated in patterned and gated
samples of graphene, such as Veselago lenses \cite{veselago} and
filters for the valley degree of freedom. \cite{valleyvalve} These
effects already occur in simple, rectangular graphitic samples,
so-called nanoribbons, which have been studied in great detail in
the past.
\cite{Fujita,Nakada,Wakabayashi,Miyamoto,brey,Ezawa,Sasaki,guinea2,guinea3}
With few exceptions, however, theoretical investigations of
electronic transport have concentrated on all-graphitic
structures. In experiments, the ultimate electronic contacts are
metallic (for illustration see, e.g., Ref. \onlinecite{Huard}).
Two recent works \cite{schomerus,blanter} have addressed the
coupling of graphene to normal-conducting electrodes, in each case
considering normal-conductor/graphene/normal-conductor (NGN)
junctions with an armchair ribbon and zigzag interfaces, as shown
Fig.\ \ref{fig1}(a). In Ref.\ \onlinecite{schomerus}, the
graphitic part was fixed at the value of charge-neutral graphene,
while the on-site potential
 in the leads was changed (the results were then
compared to the results for a set-up in which the leads are also
graphitic \cite{tworzydlo}). In Ref.\ \onlinecite{blanter}, the
on-site potential in all three regions was changed simultaneously
(the resulting finite charge-carrier density in the graphitic part
greatly enhances the conductance of the junction).

In this paper we systematically investigate the dependence of the
electronic transport through NGN junctions on independent on-site
potentials in the leads and in the graphitic region. Since the
transport at finite charge-carrier density is anisotropic and
depends on the details of the normal electrodes, we also consider
the case of zigzag ribbons with armchair interfaces [Figs.\
\ref{fig1}(c, d)], and the case of real-space leads [Figs.\
\ref{fig1}(b, d)]. We also investigate how the conductance depends
on bulk disorder, boundary roughness and interface imperfection.

Our results entail that a ballistic NGN junction can be insulating
even when the charge-carrier densities in the leads and in the
graphitic region are both finite. Conceptually, this effect can be
seen as the counterpart of the celebrated minimal, non-vanishing
conductivity exhibited by a graphene sheet at the point of
nominally vanishing charger-carrier density.
\cite{Novoselov2005,Zhang2005,ludwig,shon,guinea1,tworzydlo,Nomura,aleiner,altland,ryu,Ostrovsky,DasSarma}

We identify a simple mechanism for this insulating behavior at
finite charge-carrier density, which originates in the mismatch of
propagating modes in the normal and graphitic parts of the system.
In the transport across a ballistic interface, the transverse
momentum is conserved (modulo reciprocal lattice vectors), and the
conductance probes whether there are propagating modes with the
same transverse momentum on both sides of the interface. The
conductance thresholds, hence, are intimately related to the band
structure in the normal and graphitic parts of the junction, which
restricts the transverse momenta of propagating modes.
Consequently, the conductance thresholds could be used to deliver
information on the band structure of graphene in absolute ${\bf
k}$-space if the band structure in the leads is known. Our
numerical computations show that the insulating behavior persists
for moderate levels of bulk, edge or interface disorder, and is
only destroyed for a very rough interface. The practicality of
fabricating sufficiently clean graphene ribbons has been
demonstrated in recent experiments. \cite{kimnanoribbon} In other
mesoscopic systems, the fabrication and characterization of clean
ballistic interfaces has reached a high level of sophistication.
\cite{cleancontacts}

The paper is organized as follows. Section \ref{sec2} provides
background information on the tight-binding models used to model
the NGN junctions, and on the propagating and evanescent modes in
the normal and graphitic parts of the system. Section \ref{sec3}
presents numerically results for clean junctions and identifies
regions of insulating behavior at finite charge-carrier density.
Analytical results are given in Section \ref{sec4}. We start with
an exact calculation of the conductance for the case of armchair
ribbons with zigzag interfaces and lattice-matched leads, shown in
Fig.\ \ref{fig1}(a). The results allow us to identify the simple
mechanism for insulating behavior described above, which can be
carried over to the other geometries in Fig.\ \ref{fig1}. In
Section \ref{sec5} we discuss the effects of bulk, edge and
interface disorder, as well as the effect of mode mixing at
armchair interfaces. Conclusions are presented in Section
\ref{sec6}. Appendices \ref{appendix:transervsek} (on
transverse-momentum quantization) and \ref{appendix:gammarg} (on
the modelling of a ballistic interface to a real-space lead) give
some additional theoretical background on features of the
tight-binding model used in the numerical computations.

\afterpage{\clearpage}

\section{Theoretical Background}
\label{sec2}

In this section we provide the theoretical background for the
analytic calculations and numerical computations of the
conductance of NGN junctions, which are based on tight-binding
models and the Landauer conductance formula.

\subsection{Model Hamiltonian}

Tight-binding models of NGN junctions are shown in Fig.\
\ref{fig1}. The tight-binding Hamiltonian is given by
\begin{equation}
H=\sum_{i}V_ic_i^\dagger c_i-\sum_{\langle
ij\rangle}\gamma_{ij}c_i^\dagger c_j,
\label{hamiltonian}
\end{equation}
where $c_i$ is a fermionic annihilation operator acting on lattice
site $i$, $\langle ij\rangle$ denotes pairs of nearest neighbors,
and the hopping matrix elements $\gamma_{ij}$ as well as the
on-site potential energies $V_i$  take values as appropriate for
the region in question.

The graphitic region is modelled by sites on a honeycomb lattice
with lattice constant $a^{(G)}$, hopping constant $\gamma^{(G)}$
and on-site potential $V^{(G)}$. The normal regions $N^{(r)}$
($r=1,2$, corresponding to the left and right leads respectively)
are modelled as sites on a square lattice with lattice constant
$a^{(r)}$, hopping constant $\gamma^{(r)}$ and on-site potential
$V^{(r)}$.

In order to form an NGN junction, a graphitic region of length
${\cal L}$ is matched to the normal regions along graphitic zigzag
[Figs.\ \ref{fig1}(a,~b)] or armchair [Figs.\ \ref{fig1}(c,~d)]
interfaces of width $\mathcal{W}$. Two types of  matching are
considered. Figures \ref{fig1}(a,~c) show lattice-matched leads,
where the lattice constant is related to the lattice constant in
graphene by $a^{(r)}=\sqrt{3}a^{(G)}$ and $a^{(r)}=a^{(G)}$,
respectively. Figures \ref{fig1}(b,~d) show real-space leads,
approximated by a finer lattice with a reduced lattice constant.
The hopping constants  across the right and left interface are
denoted by $\gamma^{(1G)}$ and $\gamma^{(2G)}$, respectively.

\subsection{Landauer conductance formula}

For small bias voltages the phase-coherent conductance of a
mesoscopic structure is given by the Landauer formula
\begin{equation}
g=(2e^2/h)\,{\rm tr}\, t^\dagger t, \label{landauer}
\end{equation} where
$t$ is the transmission matrix with elements $t_{nm}$. The mode
index $n$ refers to incoming propagating modes in the left lead,
while the mode index $m$ refers to outgoing propagating modes in
the right lead. The interface couples these modes to the
propagating and evanescent modes in the graphitic scattering
region. The remainder of this background section compares the
properties of these modes in the normal and graphitic parts of the
NGN junctions.

\subsection{Dispersion relations}

The properties of the modes in the normal and graphitic regions
follow from the dispersion relations, which relate the wave number
${\bf k}=(k_x,k_y)$ to the energy  $E({\bf k})$ of Bloch waves and
reflect the symmetry properties of the underlying lattice
structure.

The unit cell of the hexagonal lattice contains two inequivalent
sites A and B with different orientation of the bonds. The unit
cell of the square lattice contains only a single site, which we
denote by S. We use the symbols $\psi_A(x,y)$, $\psi_B(x,y)$, and
$\psi_S(x,y)$ to denote the amplitudes of the  wavefunction on
each site, where $x,y$ are the coordinates of the center of the
unit cell. We assume that one unit cell is centered at the origin
$x=y=0$.

\begin{figure}[b]
\includegraphics[width=\columnwidth]{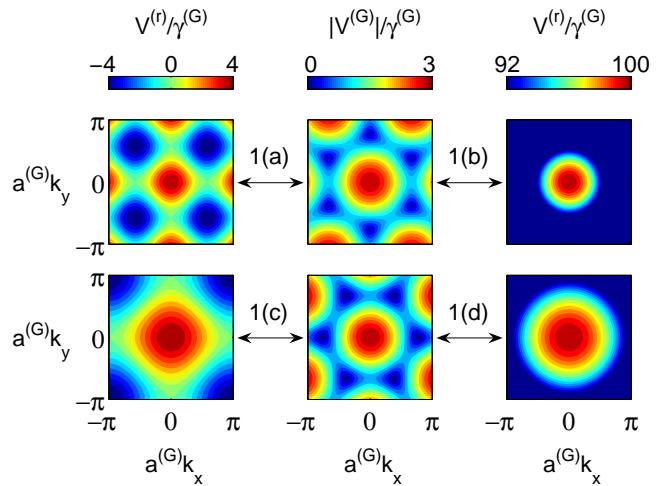}
\caption{\label{fig2}  (Color) Dependence of the Fermi lines on
the on-site potential $V^{(G)}$ or $V^{(r)}$ in the Brillouin
zones of a hexagonal lattice with hopping constant $\gamma^{(G)}$
(middle panels), a lattice-matched square lattice with hopping
constant $\gamma^{(r)}=\gamma^{(G)}$ (left panels) and a real
space lattice with lattice reduction factor $q=5$ and hopping
constant $\gamma^{(r)}=q^2\gamma^{(G)}$ (right panels). In the
upper panels, the lattice constants are related as in the NGN
junctions with zigzag interfaces, shown in Fig.\ \ref{fig1}(a, b).
In the lower panels, the lattice constants are related as in the
NGN junctions with armchair interfaces, shown in Fig.\
\ref{fig1}(c, d). The arrows indicate how the Brillouin zones have
to be matched to form the four different NGN junctions in Fig.\
\ref{fig1}. }
\end{figure}

The square lattice supports Bloch waves
\begin{equation}
\psi_S(x,y)=\psi_S(0,0) e^{ik_x x+i k_y y}
 \end{equation}
 with
dispersion relation
\begin{equation}
\label{eq:sldisp}
E=V^{(r)}-2\gamma^{(r)}[\cos(a^{(r)}k_y)+\cos(a^{(r)}k_x)].
\end{equation}
In the continuum limit $a^{(r)}\to 0$ at fixed
$\gamma^{(r)}(a^{(r)})^2\equiv\frac{\hbar^2}{2m}$, one recovers
the parabolic dispersion relation
\begin{equation}
\label{eq:parabolic} E=V^{(r)}-4\gamma^{(r)}+\frac{\hbar^2}{2m}(k_x^2+k_y^2)\,.
\end{equation}

The hexagonal lattice supports Bloch waves of the form
$(\psi_A(x,y),\psi_B(x,y))= (\psi_A(0,0),\psi_B(0,0))e^{ik_x x+i
k_y y}$. For the zigzag orientation of the interface, the
amplitudes on the A and B sites are related via
$\psi_B(0,0)=\eta\frac{f}{|f|}\psi_A(0,0)$, where the function
\begin{equation}\label{eq:f}
f(k_x,k_y)=\gamma^{(G)}[1+2e^{i3k_x a^{(G)}/2}\cos({\sqrt{3}k_y
a^{(G)} /2})]
\end{equation}
also delivers the  graphene dispersion relation via
$E=V^{(G)}-\eta|f|$. The index $\eta=\pm 1$ distinguishes the two
branches of the dispersion relation. For the armchair interface,
the graphene lattice is rotated by 90$^\circ$. The amplitudes
$\psi_B(0,0)=\eta\frac{\tilde f}{|\tilde f|}\psi_A(0,0)$ and
dispersion relation $E=V^{(G)}-\eta|\tilde f|$ are then determined
by the function \begin{equation} \tilde
f(k_x,k_y)=\gamma^{(G)}[1+2e^{i3k_y a^{(G)}/2}\cos({\sqrt{3}k_x
a^{(G)} /2})].  \label{eq:tildef}
\end{equation}

The quantization of the transverse momentum $k_y$ in a wire
geometry is discussed in Appendix \ref{appendix:transervsek}.

\subsection{Mode characterization}

Whether a mode with given transverse momentum is propagating or
evanescent is determined by the dispersion relation of the region
in question. For a given transverse momentum, the dispersion
relation delivers the longitudinal wavenumber $k_x$ as a function
of energy and on-site potential. A mode is propagating when $k_x$
is real and evanescent when $k_x$ is complex.

Propagating modes at the Fermi energy $E_F\equiv 0$ can be
identified from the condition that the line of constant $k_y$
crosses one of the Fermi lines, which depend on the on-site
potential as shown in Fig.\ \ref{fig2}. This delivers the
following conditions for propagating modes on the various types of
lattice:
\begin{subequations}\label{eq:thresholds}
\begin{eqnarray}
&&\left|\frac{|V^{(G)}|}{2\gamma^{(G)}}-\frac{1}{2}\right|<\left|\cos\left(\frac{\sqrt{3}}{2}k_y
a^{(G)}\right)
\right|<\left|\frac{|V^{(G)}|}{2\gamma^{(G)}}+\frac{1}{2}\right|
\nonumber\\
&& \mbox{(hexagonal lattice with zigzag interfaces)},
\\[.3cm]
&&1-\frac{{V^{(G)}}^2}{{\gamma^{(G)}}^2}<\left|\cos\left(\frac{3}{2}k_y
a^{(G)}\right)\right|^2 \quad \mbox{for }|V^{(G)}|<\gamma^{(G)}\nonumber\\
&&\frac{{V^{(G)}}^2}{4{\gamma^{(G)}}^2}-\frac{5}{4}<\left|\cos\left(\frac{3}{2}k_y
a^{(G)}\right)\right| \quad \mbox{for }
|V^{(G)}|>\sqrt{5}\gamma^{(G)}
\nonumber\\
&& \mbox{(hexagonal lattice with armchair interfaces)},
\\[.3cm]
&&\left|\frac{V^{(r)}}{2\gamma^{(r)}}-\cos\left(k_y a^{(r)}\right)
\right|<1
\nonumber\\
&& \mbox{(square lattice)}.
\end{eqnarray}
\end{subequations}
For each type of lattice, these conditions deliver the range of
transverse momentum in which the modes are propagating, while in
the complementary range the modes are evanescent. The border
between these ranges defines the threshold values of transverse
momentum at which the modes change their character. In Section
\ref{sec4} we translate these thresholds into conductance
thresholds for the NGN junctions.

\section{Numerical results for clean NGN junctions}
\label{sec3}

\subsection{Method and parameters}

\label{sec:dec} In order to obtain an immediate insight into the
gate-voltage dependence of the conductance of the NGN junctions of
Fig.\ \ref{fig1} we first present numerical results for clean
systems.

The transmission coefficients $t_{nm}$ are computed  employing an
efficient decimation scheme. \cite{decimation} In this scheme, the
uncoupled Hamiltonians of the leads and graphene are reduced to
effective Hamiltonians (self-energies) at the interfaces. For
square lattice leads the self-energies are known analytically.
\cite{datta}  In the graphitic region, the renormalisation
procedure is performed iteratively site-by-site employing
Gauss-Jordan elimination. The Dyson equation is then used to
determine the surface Green function of the coupled system.
Finally, the transmission coefficients follow from the Fisher-Lee
relation. \cite{fisherlee}

In our numerical computations, the graphitic region has width
$\mathcal{W}=175\,a^{(G)}$ and length $\mathcal{L}=150\,a^{(G)}$.
In the case of lattice-matched leads, panels (a) and (c) in
Fig.~\ref{fig1}, the hopping constants in the leads and across the
interfaces are all taken to be identical to the hopping constant
in graphene, $\gamma^{(r)}=\gamma^{(rG)}=\gamma^{(G)}$,
corresponding to a ballistic interface (without a tunnel barrier).
In order to model real-space leads, panels (b) and (d) in
Fig.~\ref{fig1}, the  lattice constant is reduced by a factor
$q=5$ as compared to the lattice-matched leads. The hopping
constant $\gamma^{(r)}=q^2\gamma^{(G)}$ in the leads is chosen to
preserve the effective mass $m=\hbar^2/[2\gamma^{(r)}{a^{(r)}}^2]$
in the parabolic region of the dispersion relation at the bottom
of the  band. The value $\gamma^{(rG)}=7.861\gamma^{(G)}$ for the
interface hopping term is again chosen to model a ballistic
interface (for a derivation of this value see Appendix
\ref{appendix:gammarg}).

\subsection{Results}

\begin{figure}[t]
\includegraphics[width=\columnwidth]{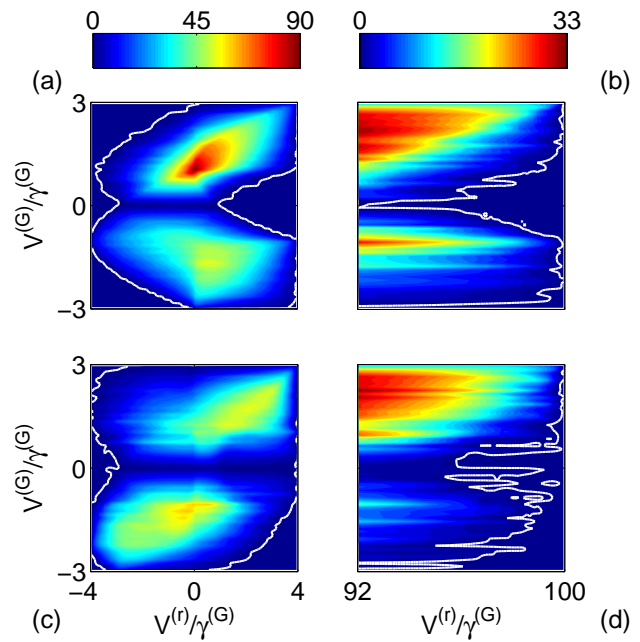}
\caption{(Color) Dependence of the conductance $g$ (in units of
the quantum of conductance $g_0=2e^2/h$) on the on-site potentials
$V^{(G})$ in graphene and $V^{(r})$ in the leads. The conductance
is color-coded as indicated in the scale bar, and the white
contour denotes the conductance thresholds where
$g=\frac{g_0}{2}$. The panels (a-d) correspond to the NGN junction
shown in Fig. \ref{fig1}(a-d). The width and length of the
graphene region were chosen to be $\mathcal{W}=175\,a^{(G)}$ and
$\mathcal{L}=150\,a^{(G)}$. Further parameters for the plots are
described in Section \ref{sec:dec}. \label{fig3}}
\end{figure}

Figures~\ref{fig3}(a-d)  show the gate voltage dependence of the
conductance of clean NGN junctions, where each panel corresponds
to one of the configurations in Fig.~\ref{fig1}(a-d). The results
are presented in a color scale where red corresponds to a large
conductance, while blue corresponds to a low conductance. The unit
of conductance is the conductance quantum $g_0=(2e^2/h)$. Regions
of low and high conductance are separated by a white contour at
$g=\frac{1}{2}g_0$. The on-site potential $V^{(G)}$ in the
graphitic part is varied independently of the on-site potential
$V^{(1)}=V^{(2)}$ in the leads. These energies are measured in
units of the hopping constant in graphene, $\gamma^{(G)}$. For
lattice-matched leads, the on-site potentials are varied over the
complete bandwidth of the dispersion relation in the graphitic
region ($-3\,\gamma^{(G)}<V^{(G)}<3\,\gamma^{(G)}$) and in the
leads ($-4\,\gamma^{(G)}<V^{(r)}<4\,\gamma^{(G)}$). For real-space
leads the on-site potential is restricted to the range
$-4\,\gamma^{(G)}+4q^2\gamma^{(G)}<V^{(r)}<4\,\gamma^{(G)}+4q^2\gamma^{(G)}$
in the parabolic region at the bottom of the square-lattice
dispersion relation, where $q=5$ is the lattice-constant reduction
factor introduced above.

The results  in Fig.~\ref{fig3} show a highly systematic
dependence of the conductance on the on-site potentials. The
conductance is small for $V^{(G)}=0$, the region of minimal
conductivity at zero charge-carrier density discussed in previous
transport studies of graphene.
\cite{Novoselov2005,Zhang2005,ludwig,shon,guinea1,tworzydlo,Nomura,aleiner,altland,ryu,Ostrovsky,DasSarma}
However, we also find regions of very small conductance where the
charge-carrier density in graphene and in the leads is finite.
Conditions for this insulating behavior are determined in Section
\ref{sec4}. The general raggedness of the contours of constant
conductance is a common feature in graphene transport studies, and
can be associated to Fabry-Perot resonances.
\cite{guinea2,guinea3,blanter}

For lattice-matched  leads with zigzag interfaces
[Fig.~\ref{fig3}(a)], we observe an approximate mirror symmetry of
the conductance for $(V^{(r)},V^{(G)})\to (V^{(r)},-V^{(G)})$.
This symmetry is most pronounced for the regions of low
conductance, delimited by the white threshold contour. On the
other hand, the maximal conductance for positive $V^{(G)}$ is much
larger than for negative $V^{(G)}$. These maxima are found at
$V^{(r)}\approx 0$ and $V^{(G)}\approx \pm\gamma^{(G)}$.

For lattice-matched leads  with armchair interfaces
[Fig.~\ref{fig3}(c)], the mirror symmetry $(V^{(r)},V^{(G)})\to
(V^{(r)},-V^{(G)})$ is only observed for the region of low
conductance with $V^{(r)}\approx-4\,\gamma^{(G)}$, close to the
top of the square-lattice dispersion relation. The region of high
conductance obeys an approximate symmetry when both on-site
potentials are inverted, $(V^{(r)},V^{(G)})\to
(-V^{(r)},-V^{(G)})$.

For real-space leads [Fig.~\ref{fig3}(b,~d)], the general features
of the conductance are inherited from the behavior for the
lattice-matched leads in the parabolic region at the bottom of the
square-lattice dispersion relation. In this region, the
conductance in general increases for increasing charge-carrier
density in the leads (corresponding to smaller values of
$V^{(r)}$), and large values of the conductance are predominantly
found for positive $V^{(G)}$. Despite being more ragged, the
threshold contours have a similar general trend as in the
lattice-matched case. For zigzag interfaces, the region of
insulating behavior obeys the approximate mirror symmetry, while
this is not the case for armchair interfaces.

\section{Analytical results}
\label{sec4}

Most of the conductance thresholds observed in the numerical
computations, Fig.\ \ref{fig3}, can be explained via a simple
mechanism, based on the mismatch of propagating modes on both
sides of an NG interface. We start our considerations with the
exact calculation of the conductance of ballistic NGN junctions
with zigzag interfaces and latticed-matched leads [Fig.\
\ref{fig1}(a)]. The calculation shows that in this case, the
insulating regions correspond to conditions where the propagating
modes in the normal leads only couple to evanescent modes in the
graphitic scattering region. This observation is then carried over
as a criterion to calculate conductance thresholds for the other
three types of NGN junctions.

\subsection{Conductance for NGN junctions with zigzag interfaces and lattice-matched leads}

For ballistic NGN junctions with zigzag interfaces and
latticed-matched leads [see Fig.\ \ref{fig1}(a)] the conductance
can be calculated analytically via a wave matching procedure. The
calculation succeeds because for the zigzag configuration the
hard-wall  boundary conditions in the N and G parts select the
{\em same} transverse wave components, Eqs.\
(\ref{ky1},\ref{ky2}). Hence no mode mixing occurs at a clean
zigzag interface. The transmission matrix becomes diagonal, and
the wave matching for each fixed transverse-mode profile $n$
reduces to a one-dimensional problem.

A derivation of the matching conditions for the present geometry
has been given in Refs.\ \onlinecite{schomerus,blanter}. The
wavefunction in the square leads, $\psi_S(x,y)=\psi_S(0,0)e^{ik_xx
+ ik_yy}$, has to be matched with the wavefunction in the
graphitic region, $(\psi_A(x,y) , \psi_B(x,y)) = (\psi_A(0,0) ,
\psi_B(0,0))e^{ik_xx + ik_yy}$ at the interfaces of the NGN
junction (located at $x=0$ and $x=L$), subject to the boundary
conditions
\begin{eqnarray}
\nonumber \gamma^{(G)}\psi_A(0,0) &=& \gamma^{(1G)}\psi_S(0,0)\,, \\
\nonumber \gamma^{(1G)}\psi_B(0,0) &=& \gamma^{(1)}\psi_S(a^{(1)},0)\,, \\
\nonumber \gamma^{(2G)}\psi_A(L,0) &=& \gamma^{(2)}\psi_S(L-a^{(2)},0)\,, \\
\gamma^{(2)}\psi_B(L,0) &=& \gamma^{(2G)}\psi_S(L,0)\,.
\end{eqnarray}

In Refs.\ \onlinecite{schomerus,blanter} these equations have been
solved for the cases $V^{(G)}=0$, $V^{(1)}=V^{(2)}$ of
charge-neutral graphene and $V^{(1)}=V^{(2)}=V^{(G)}$ for
uniformly gated junctions, respectively. For the general case of
independent on-site potentials and coupling constants we find
\begin{widetext}
\begin{eqnarray}
t_{nn}=&&-4iC\sin\left(
\frac{3}{2}a^{(G)}k_x^{(G)}\right)\sin\left(\sqrt{3}a^{(G)}k_x^{(1)}\right)\sin\left(\sqrt{3}a^{(G)}k_x^{(2)}\right)
\times \nonumber \\
&& \times \Bigg\{ \left(
\Gamma_1+\frac{1}{\Gamma_1\lambda^{(1)}\lambda^{(2)}}-
\frac{V^{(G)}\Gamma_2}{\gamma^{(G)}\lambda^{(1)}}-
\frac{V^{(G)}}{\gamma^{(G)}\Gamma_2\lambda^{(2)}}
  \right)\sin\left( {\cal L}k_x^{(G)}\right)+{}\nonumber \\
&&
  {}+2C\Gamma_1\sin\left[
\left({\cal L}-\frac{3}{2}a^{(G)}\right)k_x^{(G)}\right]
+\frac{2C}{\Gamma_1\lambda^{(1)}\lambda^{(2)}}\sin\left[
\left({\cal L}+\frac{3}{2}a^{(G)}\right)k_x^{(G)}\right]
  \Bigg\}^{-1}
  ,
\label{result}
\end{eqnarray}
where the $n$ dependence of the above expression is implicit in both $k_{x}$ and $k_{y}$, and
\begin{eqnarray}
   &&C=\cos\left(\frac{\sqrt{3}}{2}a^{(G)}k_y\right),
\qquad \lambda^{(r)}=e^{i\sqrt{3}a^{(G)}k_x^{(r)}},\qquad
  \Gamma_1=\frac{\gamma^{(1G)}\gamma^{(2G)}}{\sqrt{\gamma^{(1)}\gamma^{(2)}}\gamma^{(G)}}
,\qquad
\Gamma_2=\frac{\gamma^{(2G)}}{\gamma^{(1G)}}\frac{\sqrt{\gamma^{(1)}}}{\sqrt{\gamma^{(2)}}}.
\end{eqnarray}
\end{widetext}
The conductance of the junction follows from the Landauer formula
(\ref{landauer}). In the limit of  $V^{(G)}=0$ and
$V^{(1)}=V^{(2)}$ one recovers the result of
Ref.~\onlinecite{schomerus}, while for $V^{(1)}=V^{(2)}=V^{(G)}$
the result of Ref.~\onlinecite{blanter} is obtained. We confirmed
that the values of conductance obtained from Eq.~(\ref{result})
for general combinations of the on-site potentials are in
numerical agreement with the results obtained in the previous
section [Fig.\ \ref{fig3}(a)].

\subsection{Conductance thresholds in large ballistic
NGN  junctions due to mode mismatch}

\begin{figure}[b]
\includegraphics[width=\columnwidth]{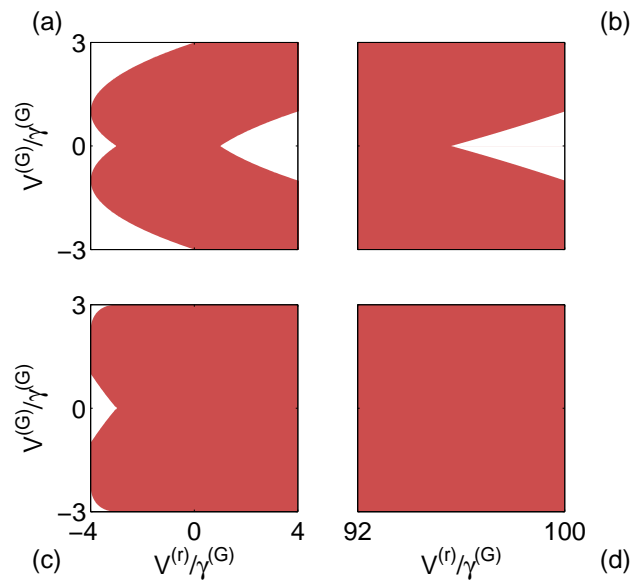}
\caption{(Color online) For each of the NGN junction shown in
Figs.\ \ref{fig1}(a-d), the white regions denote conditions for
insulating behavior due to the mismatch of propagating modes. The
boundary of the regions are the conductance thresholds derived in
Eq.\ (\ref{eq:thresh1a}) (for panel a), Eq.\ (\ref{eq:thresh1b})
(for panel b), and  Eq.\ (\ref{eq:thresh1c}) (for panel c). No
regions of mismatching modes occur in panel (d). Comparison to
Figs.\ \ref{fig3}(b, d) shows that the conductance can be small
even beyond the mode-mismatch mechanism. \label{fig4}}
\end{figure}

For a long graphitic region, with ${\cal L}\gg a^{(G)}$, a
transmission coefficient (\ref{result}) tends to zero when the
propagating mode in the lead couples to an evanescent mode in the
graphitic region (i.e., if $\mathrm{Im}\,k_{x}^{(G)} \neq 0$). The
conductance (\ref{landauer}) of the junction is small if this is
the case for all transmission coefficients. A small conductance,
hence, does not necessitate that all the modes in the graphitic
region are evanescent -- it suffices that the propagating modes in
the graphitic region do not couple to the propagating modes in the
leads. Consequently, the conductance can be small even when the
charge-carrier density both in the graphitic region as well as in
the normal leads is finite. We now apply this mode-mismatch
mechanism to calculate conduction thresholds of long and wide
ballistic NGN junctions with clean interfaces, covering all of the
geometries shown in Fig.\ \ref{fig1}.

The requirement of a wide junction arises from the fact that the
mechanism described above relies on the conservation of transverse
momentum (modulo reciprocal-lattice vectors). For a clean
zigzag-interface, this conservation law is exact. As discussed in
Appendix \ref{appendix:transervsek}, for an armchair interface,
the quantized transverse momenta in the graphitic part,
 Eq.\ (\ref{kyarmchair}), differ from the quantized transverse
momenta in the normal leads, Eq.\ (\ref{ky2}). The resulting
mode-mixing for interfaces of finite width is discussed in Section
\ref{armmodemixing}. Moreover, details of the transverse-momentum
quantization in graphene ribbons depend on the chemistry of the
edges. \cite{Kawai,Barone,Son} In wide junctions the modes in N
and G are tightly spaced and can be assumed to be
quasi-continuous. In the limit ${\cal W}\to\infty$, the transverse
wave number $k_y$ is then conserved exactly, also for armchair
interfaces.

The requirement of a long, ballistic NGN junction is needed so
that we can use the assumption that the evanescent modes only give
negligible contributions to the conductance, as explained in more
detail below. (The effects of disorder are discussed in Section
\ref{sec5}.)

Under these conditions, good conduction requires values of the
on-site potentials $V^{(G)}$ and $V^{(r)}$ at which one finds
transverse momenta, possibly differing by reciprocal-lattice
vectors, for which the associated modes in N and G are both
propagating. The conductance is always small when this criterion
is not fulfilled. The threshold values of the on-site potentials
separating the regions of matching and mismatching propagating
modes can be derived from Eqs.\ (\ref{eq:thresholds}).

For the arrangement in Fig.\ \ref{fig1} (a), the conductance
threshold due to mode mismatch has two branches given by
\begin{subequations}\label{eq:thresh1a}
\begin{eqnarray}
\frac{V^{(r)}}{\gamma^{(r)}}&=&\left(\frac{|V^{(G)}|}{\gamma^{(G)}}+1\right)^2,
\\
\frac{V^{(r)}}{\gamma^{(r)}}&=&\left(\frac{|V^{(G)}|}{\gamma^{(G)}}-1\right)^2-4.
\end{eqnarray}
\end{subequations}

For a finer discretization of the square lattice as in Fig.\
\ref{fig1} (b), the first branch remains within the region of the
parabolic dispersion at the bottom of the band, while the second
branch shifts to $V^{(r)}\to -\infty$. In the continuum limit
(\ref{eq:parabolic}) of the dispersion relation, conductance
thresholds due to mode mismatch exist only for
$\frac{\sqrt{2m(4\gamma^{(r)}-V^{(r)})}}{\hbar}\sqrt{3}a^{(G)}
<2\pi/3$, and then are given by
\begin{equation} \label{eq:thresh1b}
\frac{|V^{(G)}|}{\gamma^{(G)}}=2\cos\left(\frac{\sqrt{2m(4\gamma^{(r)}-V^{(r)})}}{\hbar}\frac{\sqrt{3}}{2}a^{(G)}\right)-1.
\end{equation}
The survival of conductance thresholds in the continuum limit can
be best understood by considering the modes with $k_y\approx 0$,
which propagate in the normal leads for on-site potentials close
to the bottom of the parabolic dispersion relation. For zigzag
interfaces, these modes propagate  in the graphitic region only in
the region $|V^{(G)}|>\gamma^{(G)}$ [see Fig.\ \ref{fig2}].

For the arrangement in Fig.\ \ref{fig1} (c), care has to be taken
for the fact that the periods of the Brillouin zones of the leads
and the graphene part differ by a factor of $3/2$  in the $k_y$
direction (see the alignment of the Brillouin zones in Fig.\
\ref{fig2}). The graphitic region hence mediates the coupling of
lead modes with different transverse momentum. Propagating modes
always match up for $v\equiv\frac{V^{(r)}}{\gamma^{(r)}} \geq -3$,
while in the region $v< -3$ there are three branches of
conductance thresholds due to mode mismatch. Two of these branches
are bounded by the condition
\begin{subequations}\label{eq:thresh1c}
\begin{eqnarray}
\frac{|V^{(G)}|}{\gamma^{(G)}}&=&\sqrt{5+4\sqrt{1+\frac{9}{4}v+\frac{3}{2}v^2+\frac{1}{4}v^3}}.
\end{eqnarray}
The third branch is given by the condition
\begin{eqnarray}
\frac{|V^{(G)}|}{\gamma^{(G)}}&=&\sqrt{-\frac{9}{4}v-\frac{3}{2}v^2-\frac{1}{4}v^3}.
\end{eqnarray}
\end{subequations}

For a finer discretization of the square lattice as in Fig.\
\ref{fig1} (d), all these branches move to $V^{(r)}\to -\infty$.
This results in the absence of  conductance thresholds due to mode
mismatch in the continuum limit. This can be understood from the
observation that in graphene with armchair interfaces, one can
find propagating modes with $k_y\approx 0$ for all values of the
on-site potential $V^{(G)}$ [see again Fig.\ \ref{fig2}].

Figure \ref{fig4} shows the conductance thresholds due to mode
mismatch in the $V^{(G)}$ -- $V^{(r)}$ plane. Each panel
corresponds to one of the various types of NGN junctions shown in
Fig.\ \ref{fig1}. Comparison with Fig.\ \ref{fig3} shows that for
zigzag interfaces [panels (a) and (b)], the mode-mismatch
mechanism explains all conductance thresholds. For armchair
interfaces, the mode-mismatch mechanism explains the conductance
thresholds in the left part of panel (c), corresponding to
energies close to the top of  the  band in the leads. The
numerical results in Fig.\ \ref{fig3}(c, d) exhibit additional
thresholds in the lower-right corner of the $V^{(r)}$-$V^{(G)}$
plane, corresponding to the bottom of the conduction band in the
leads and the top of the conduction band in the graphitic part of
the system. Here the propagating modes on both sides of the
interface differ drastically in their longitudinal wavenumber (and
hence in their self-energy), which also inhibits their coupling.
\cite{schomerus} Consequently, the conductance of a ballistic NGN
junction can be small even for conditions not described by the
simple mode-mismatch mechanism.

\subsection{Sharpness of thresholds}

Above we have ignored the contribution of evanescent modes in
graphene. These modes become important for a finite system size
$L$, and determine the sharpness of the conductance thresholds.

The role of the evanescent modes is best understood by considering
the most slowly decaying modes in the graphitic region, and in
particular by investigating which of these modes still couple to
propagating modes in the leads when one enters the insulating
regime. It comes in handy that the most slowly decaying modes have
transverse wave numbers just at the threshold to where such modes
become propagating, which is determined by Eq.\
(\ref{eq:thresholds}). Inside the insulating region, not only all
the propagating modes in graphene, but also the adjacent slowly
decaying evanescent modes couple to evanescent modes in the leads
and hence do not contribute to the transport. The remaining
graphitic evanescent modes which do couple to the propagating
modes in the leads all have a finite decay constant  ${\rm Im}\,
k_x^{(G)}>\kappa$, and their total contribution to the conductance
is suppressed exponentially with $\exp(-\kappa L)$. The sharpness
of the thresholds hence increases exponentially with the system
size.

The decay constant $\kappa$ approaches zero as one approaches the
conductance thresholds. Let us assume that this is induced by
changing the on-site potential $V^{(r)}\to V^{(r),\rm thresh}$ at
fixed $V^{(G)}$, where depending on the geometry $V^{(r),\rm
thresh}$ is determined by Eq. (\ref{eq:thresh1a}),
(\ref{eq:thresh1b}), or (\ref{eq:thresh1c}). For a vanishing
charge-carrier density in graphene, $V^{(G)}=0$, the linear
dispersion relation close to the Dirac point then entails that
$\kappa\propto |V^{(r)}- V^{(r),\rm thresh}|$ increases linearly
with the distance to the threshold, while for a finite $V^{(G)}$
it increases faster, as $\kappa\propto |V^{(r)}-V^{(r),\rm
thresh}|^{1/2}$. Hence, the conductance thresholds are sharper at
a finite charge density.

It is insightful to contrast the exponential suppression of the
conductance carried by  evanescent modes in the insulating region
with their contribution inside the conductive region. In this case
the most slowly decaying modes do couple to propagating modes in
the leads. For a vanishing charge-carrier density in graphene,
$V^{(G)}=0$, the conductance carried by the evanescent modes adds
up to a contribution $\propto W/L$, which is constant and finite
at a fixed aspect ratio even when the system is very large.
\cite{tworzydlo,schomerus} For finite $V^{(G)}$, on the other
hand, their contribution is proportional to $W/L^4$ for $L\gg
a^{(G)} \gamma^{(G)}/|V^{(G)}|$ and hence decays algebraically for
increasing system size. \cite{blanter} Both expressions require
that the most slowly decaying evanescent modes in the graphitic
part couple to the propagating modes in the leads, which is not
the case in the insulating regime.

\section{Mode mixing}

\label{sec5}

\begin{figure}[tb]
\includegraphics[width=\columnwidth]{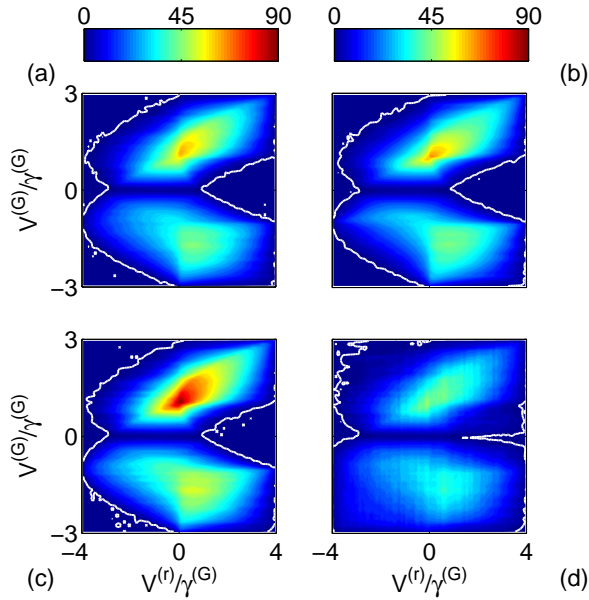}
\caption{\label{fig5} (Color) Same as Fig.\ \ref{fig3}(a), but in
the presence of bulk disorder with $u=0.2$ (panel a), surface
roughness with $f=0.3$ (panel b), and interface disorder with
$\Gamma=0.1\,\gamma^{(rG)}$ and $\Gamma=\gamma^{(rG)}$ (panels c
and d, respectively).}
\end{figure}

The derivation of conductance thresholds from the condition of
mismatching propagating modes in   Section \ref{sec4} relied on
the conservation of transverse momentum. In this section we
explore how violations of this assumption modify the threshold
conditions.

\subsection{Mode mixing by disorder}

In order to explore the effects of mode mixing by disorder, we
implement three different scattering mechanisms: short-ranged bulk
disorder and surface roughness in the graphitic region, as well as
imperfections at the NG interfaces (long-ranged bulk disorder does
not provide efficient mode-mixing). Bulk disorder is modelled via
a random on-site potential $V_i=V^{(G)}+u_i$ where the $u_i$ are
independently and identically distributed (i.i.d.) random numbers
drawn with uniform probability from an interval $[-u/2,u/2]$. For
a rough edge we randomly eliminate a fraction $f$ of the graphene
sites within a distance of $2a^{(G)}$ from the boundaries of the
system. An imperfect interface is modelled via random hopping
elements $\gamma_{ij}= \gamma^{(rG)}+\Gamma_{ij}$ for the links
crossing the interface, where the $\Gamma_{ij}$ are i.i.d. random
numbers drawn with uniform probability from an interval
$[-\Gamma/2,\Gamma/2]$.

Figure \ref{fig5} presents the results for an NGN junction with
zigzag interface and lattice-matched leads [the geometry of Fig.
\ref{fig1}(a); the results for the other geometries are similar].
Panel (a) shows the conductance for bulk disorder of strength
$u=0.2\gamma^{(G)}$. Panel (b) shows the conductance for surface
roughness with $f=0.3$, the value for which we found the strongest
effect on the conductance. In both cases, the maximal conductance
is reduced to about $2/3$ of the value found in the clean case.
This is comparable to what is found in other transport studies at
similar levels of disorder.
\cite{Munoz-Rojas,guineasurface,Areshkin,li,martinsurface} In
contrast, the threshold contours delimiting the region of low
conductance are only weakly affected by the disorder.

Figure \ref{fig5}(c) shows the conductance for an imperfect
interface $\Gamma=0.1\gamma^{(rG)}$. This moderate level of
imperfection has only a minimal effect on the regions of high and
low conductance. A noticeable change of these regions is only
induced for a rough interface, shown in Fig.\ \ref{fig5}(d), where
the fluctuations $\Gamma=\gamma^{(rG)}$ are set equal to the
average interface hopping element. At this level of imperfection,
the regions of low conductance cover a smaller part in the
$V^{(G)}-V^{(r)}$ plane. This has to be attributed to the
diffractive effects of a rough interface. It is interesting to
observe that the conductance thresholds are most robust around
$V^{(G)}=0$; especially, the conductance threshold in the region
$V^{(r)}<0$ is almost unchanged even though the interface is very
rough.

\subsection{Mode mixing at clean armchair interfaces}

\label{armmodemixing}

\begin{figure}[t]
\includegraphics[width=\columnwidth]{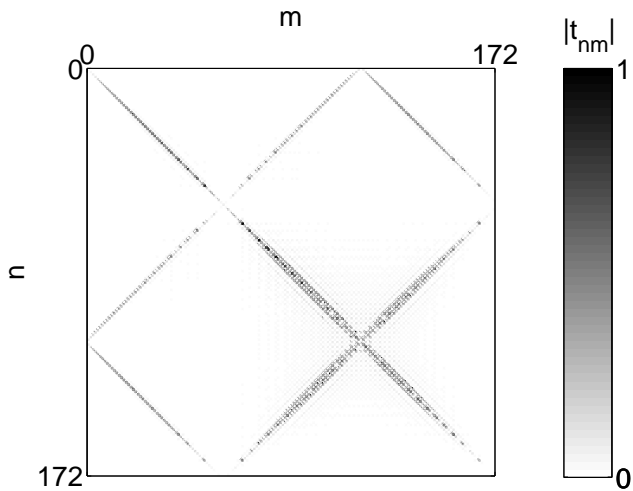}
\caption{\label{fig6} Density plot of the transmission
coefficients $|t_{nm}|$ for an NGN junction with armchair
interfaces and lattice-matched leads [the geometry of Fig.\
\ref{fig1}(c)], with $V^{(G)}=\gamma^{(G)}$, $V^{(r)}=0$ (other
parameters as in Fig.\  \ref{fig3}(c)).}
\end{figure}

As discussed in Appendix \ref{appendix:transervsek}, the quantized
transverse momenta in the graphitic part of an NGN junction with
armchair interfaces (and zigzag surface), Eq.\ (\ref{kyarmchair}),
differ from the quantized transverse momenta in the normal leads,
Eq.\ (\ref{ky2}). This results in a finite amount of mixing even
for a clean interface, which is automatically accounted for in the
numerical results of Section \ref{sec2}. Figure \ref{fig6} shows a
density plot of the modulus $|t_{nm}|$ of the transmission
amplitudes for the NGN junction in Fig.\ \ref{fig1}(c), with
parameters as for the computations in Fig.\ \ref{fig3}(c). The
on-site potentials are set to the values $V^{(G)}=\gamma^{(G)}$,
$V^{(r)}=0$, where all modes are propagating, so that the mode
mixing can be seen in transmission. The figure shows that the
transmission matrix is sparse. Each mode mixes with a small number
of modes with similar transverse momentum. The additional branches
originate from the different periodicity of the Brillouin zones in
the leads and the graphitic part, already mentioned in Section
\ref{sec4}. At each interface, the transverse momentum is only
conserved modulo reciprocal lattice vectors. The periodicity of
the Brillouin zone of the leads and the graphitic region in $k_y$
direction differs by a factor of $3/2$. For the conditions of
Fig.\ \ref{fig6}, this mediates the coupling into two additional
branches of transverse momenta in the leads.

\section{\label{sec6}Conclusions}

In this work we systematically investigated the gate-voltage
dependence of the conductance of four variants of
normal-conductor/graphene/normal-conductor (NGN) junctions,
consisting of a graphene strip which is coupled in different ways
to normal leads of identical width (c.f. Fig.\ \ref{fig1}).
Starting from exact numerical computations we identified
conditions of insulating behavior in clean junctions, which can be
encountered even when the charge-carrier density in the central
graphitic region and in the normal-conducting leads is finite.
Conceptually, this effect can be seen as the counterpart to the
celebrated minimal, finite conductivity of graphene close to the
charge-neutrality point, where the charge-carrier density
nominally vanishes.
\cite{Novoselov2005,Zhang2005,ludwig,shon,guinea1,tworzydlo,Nomura,aleiner,altland,ryu,Ostrovsky,DasSarma}

We identified a simple mechanism for the conductance thresholds at
finite charge-carrier density, namely the decoupling of the
propagating modes in the different parts of the system due to the
mismatch of their transverse momenta. Since these momenta are
determined by the dispersion relation, the conductance thresholds
could in principle be used to obtain information about the band
structure of graphene if the dispersion relation in the leads is
known. Our numerical computations show that such an analysis would
be
 robust against the effects of bulk and surface
disorder in the graphitic region, and would also tolerate a
moderate amount of imperfection of the interfaces.

  \acknowledgments

We gratefully acknowledge helpful discussions with Vladimir Fal'ko
and Edward McCann, as well as useful correspondence with Andre
Geim and Philip Kim.
  This work was supported
by the European Commission, Marie Curie Excellence Grant
MEXT-CT-2005-023778.
\appendix

\section{Transverse-momentum quantization}
\label{appendix:transervsek} In a wire geometry, the boundary
conditions select a discrete set of $W$ transverse wave numbers
$k_y$ for a given propagation or decay direction, which we
enumerate by a mode index $n=1,2,3,\ldots,W$. The details of the
transverse-momentum quantization of graphene ribbons depend on the
chemistry of the edges. \cite{Kawai,Barone,Son} In this paper, we
are mostly concerned with wide graphitic regions, where the
transverse momentum becomes quasi-continuous. When we, in the
following, give expressions for $W$ in the tight-binding model
used in the numerical simulations, it should be noted that the
dimension ${\cal W}$ refers to the width of the interface, which
is identical to the width of the square-lattice leads but differs
from the width of the graphitic region (which is wider by
$\sqrt{3}a^{(G)}$ for zigzag interfaces and by $a^{(G)}$ for
armchair interfaces).

On the square lattice, $W=1+{\cal W}/a^{(r)}$ is the number of
sites in the cross-section of the wire, and the set of quantized
transverse wave numbers is given by
\begin{equation}
k_{y}=\frac{n\pi}{{\cal W}+2a^{(r)}}.  \label{ky1}
\end{equation}
On the hexagonal lattice with zigzag interfaces (and hence
armchair boundaries), $W=1+{\cal W}/\sqrt{3}a^{(G)}$, and the set
of quantized transverse wave numbers is given by
\begin{equation}
k_y=\frac{n\pi}{{\cal W}+2\sqrt{3}a^{(G)}}. \label{ky2}
\end{equation}
Both sets of quantized transverse wave numbers become identical
when the lattices are matched commensurably as shown in Fig.\
\ref{fig1}(a), where $a^{(r)}=\sqrt{3}a^{(G)}$.

More complicated is the case of the hexagonal lattice with
armchair interface (which has zigzag edges), shown in Fig.\
\ref{fig1}(b). At the upper (lower) edge, hard-wall boundary
conditions translate into a vanishing amplitude on the A (B) sites
in the first unit cell beyond the wire boundary. Since the
amplitudes on these sites carry a relative phase which depends on
the propagation direction, the quantized transverse wave numbers
are determined by a transcendental equation,
\begin{widetext}
\begin{equation}\label{kyarmchair}
e^{2i({\cal W}+7a^{(G)})k_y}=
\frac{\sqrt{{V^{(G)}}^2/{\gamma^{(G)}}^2-\sin^2(3k_ya^{(G)}/2)}+i\sin(3k_ya^{(G)}/2)
}{\sqrt{{V^{(G)}}^2/{\gamma^{(G)}}^2-\sin^2(3k_ya^{(G)}/2)}-i\sin(3k_ya^{(G)}/2)}.
\end{equation}
\end{widetext} For $|V^{(G)}|<\gamma^{(G)}$ and close to the K
points, this equation reduces to the condition derived from the
Dirac equation given in  Ref.\ \onlinecite{brey}. In general, Eq.\
(\ref{kyarmchair}) has $W=(4/3)+2{\cal W}/3a^{(G)}$ independent
solutions. For $|V^{(G)}|<\gamma^{(G)}$, this includes a number of
edge states with $\mathrm{Im}\, k_y\neq 0$. For wide interfaces
($W\gg 1$), the real-valued transverse wave numbers are almost
uniformly spaced, but do not coincide with the transverse wave
numbers of the lattice-matched square lattice with $a^{(r)} =
a^{(G)}$, shown in Fig.\ \ref{fig1}(b).

\section{Interface hopping constant for a transparent interface with a real-space lattice}
\label{appendix:gammarg}

\begin{figure}[tb]
\includegraphics[width=\columnwidth]{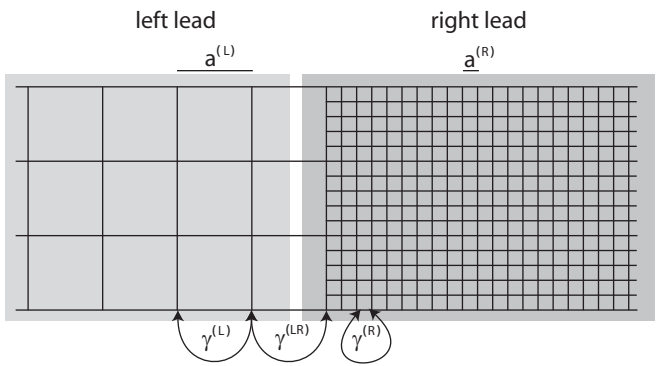}
\caption{\label{fig7} Interface between two commensurably matched
square-lattice leads of different lattice constant
(lattice-constant reduction factor $q=5$).}
\end{figure}

In this appendix we describe how to model a transparent interface
between a tight-binding lattice and a real-space lattice, as shown
in Fig.\ \ref{fig1}(b,d). A simpler variant of such an interface,
well suited for analytical calculations, is the interface of two
commensurably matched square-lattice leads with different lattice
constant, shown in Fig.\ \ref{fig7}. The lattice constants in the
left and right lead are related by an integer-valued reduction
constant $q$ so that $a^{(R)} = a^{(L)}/q$; the continuum limit of
the right lead is approached for $q\to\infty$. This simple
arrangement is representative for couplings of real space leads to
other tight-binding lattices, since only the sites adjacent to the
interface enter the subsequent considerations.

The hopping constants in the left and right lead  are denoted by
$\gamma^{(L)}$ and $\gamma^{(R)}$, respectively. In our numerical
computations we further made the assumption that these constants
are related by the requirement of an identical effective mass
$m=\hbar^2/[2\gamma^{(R)}(a^{(R)})^2]=\hbar^2/[2\gamma^{(L)}(a^{(L)})^2]$.
The hopping constants in the two leads are then related by
$\gamma^{(R)}=q^2\gamma^{(L)}$.

 Our goal is to determine the inter-lead hopping
constant, $\gamma^{(LR)}$, so that the lead is transparent for
energies in the parabolic region of the dispersion relation at the
bottom of the bands of both leads.

We choose a coordinate system where  the right lead starts at
$x=0$ and the origin accommodates one of the lattice sites that is
linked to the left lead. Now consider a particle arriving from the
left lead at a fixed energy $E$ and transverse wave number $k_y$,
which are both conserved under reflection from the interface. The
wave function
\begin{equation}\label{eq:psiL}
\psi^{(L)}(x,y)=A e^{ik_x^{(L)} x+ik_y y}+B e^{-ik_x^{(L)} x+ik_y
y}
\end{equation}
then describes  the superposition of the incoming and reflected
particle, where the longitudinal wave number $k_x^{(L)}$ is fixed
by the dispersion relation (\ref{eq:sldisp}).

Upon crossing the interface, the transverse momentum is conserved
modulo reciprocal lattice vectors. Since the Brillouin zone in the
right lead is larger by a factor $q$, one couples into $q$
inequivalent modes with wavenumber $k_{y,p}=k_y+2\pi p/a^{(L)}$,
$p=0,1,\ldots,q-1$. For energies close to the bottom of the bands
in both leads, only the mode with $p=0$ is propagating while the
others are evanescent. We denote the longitudinal wavenumber of
the propagating mode by $k_x^{(R)}$, and the decay constant of the
evanescent modes by  $\kappa_p$, $p=1,2,\ldots,q-1$. The wave
function in the right lead is hence given by
\begin{equation}\label{eq:psiR}
\psi^{(R)}(x,y)=C e^{ik_y y}\left[e^{ik_x^{(R)} x}+
\sum_{p=1}^{q-1}c_p e^{-\kappa_p x+2\pi i p y/a^{(L)}}\right].
\end{equation}
The relative amplitudes $c_p$ of the evanescent modes are fixed by
the boundary condition on those sites at $x=0$ which have no link
to left lead. Since the wavefunction (\ref{eq:psiR}) would fulfil
the Schr{\"o}dinger equation when the right lead would be
continued beyond the interface, this boundary condition can be
formally expressed as $\psi^{(R)}(-a^{(R)},y)=0$, where
$y=a^{(R)}p'$, with $p'=1,2,\ldots,q-1$, is the transverse
coordinate of inequivalent disconnected sites. These boundary
conditions are fulfilled when the amplitudes take the value
$c_p=e^{-\kappa_p a^{(R)}-ik_x^{(R)} a^{(R)}} $.

The remaining amplitudes $A$, $B$, and $C$ are now obtained from
the boundary conditions of the sites on both leads which are
linked to the other lead. Using again the fact that the
wavefunctions (\ref{eq:psiL}) and (\ref{eq:psiR}) both fulfil the
Schr{\"o}dinger equation when the leads would be extended across
the interface, these conditions can be written as
\begin{eqnarray}
&&\gamma^{(RL)}\psi^{(R)}(0,0)=\gamma^{(L)}\psi^{(L)}(0,0),\\
&&\gamma^{(R)}\psi^{(R)}(-a^{(R)},0)=\gamma^{(RL)}\psi^{(L)}(-a^{(L)},0).
\end{eqnarray}
The reflection amplitude $r=B/A$ then follows as
\begin{equation}
r=-\frac{\tilde q
{\gamma^{(RL)}}^2e^{-ik_x^{(L)}a^{(L)}}-q\gamma^{(L)}\gamma^{(R)}
e^{-ik_x^{(R)}a^{(R)}}}{\tilde q
{\gamma^{(RL)}}^2e^{ik_x^{(L)}a^{(L)}} -q\gamma^{(L)}\gamma^{(R)}
e^{-ik_x^{(R)}a^{(R)}}},
\end{equation}
where
\begin{equation}
\tilde q =1+\sum_{p=1}^{q-1}c_p
\end{equation}

 At the bottom
of the band, we can further assume $k_y, k_x^{(R,L)}\approx 0$,
and the decay constants approach the value
$a^{(R)}\kappa_p=\mathrm{arcosh}\,[2-\cos(2\pi p/q)]$. The
condition $r=0$ for a transparent interface then gives the desired
expression for the coupling constant, which after some algebraic
manipulations can be written as
\begin{eqnarray}
&& \gamma^{(LR)} = \sqrt{\frac{\gamma^{(L)} \gamma^{(R)}}{2-s}},
 \label{eq:gammalr} \\
&&s=\frac{1}{q}\sum_{p=1}^{q-1}\sqrt{3-4\cos\frac{2\pi p}{q}
+\cos^2\frac{2\pi p}{q}} \,.\nonumber
\end{eqnarray}

The value $\gamma^{(LR)}=7.861\gamma^{(L)}$ used in the numerical
computations finally follows when Eq.\ (\ref{eq:gammalr}) is
evaluated for $\gamma^R=q^2 \gamma^L$ and a  lattice-constant
reduction factor $q=5$.

\end{document}